\documentclass[%
 reprint,
superscriptaddress,
 amsmath,amssymb,
 aps,
prb,
]{revtex4-2}

\usepackage{graphicx}
\usepackage{dcolumn}
\usepackage{bm}
\usepackage{xcolor}
\usepackage[utf8]{inputenc}
\usepackage{amsmath}
\usepackage{amssymb}
\usepackage{graphicx}
\usepackage{gensymb}
\usepackage{graphics}
\usepackage[T1]{fontenc}
\usepackage{soul}
\usepackage[para,online,flushleft]{threeparttable}
\PassOptionsToPackage{hyphens}{url}\usepackage{hyperref}

\begin{document}

\preprint{APS/123-QED}

\title{Manuscript Title:\\Twist-tunable spin control in twisted bilayer bismuthene}

\author{Ludovica Zullo}
\email[]{\\ludovica.zullo@unitn.it}
\affiliation{Department of Physics, University of Trento, Via Sommarive 14, 38123 Povo, Italy}
\affiliation{Sorbonne Universit\'e, CNRS, Institut des Nanosciences de Paris, UMR7588, F-75252 Paris, France}
\affiliation{Dipartimento di Fisica ``E. Pancini'', Universit\`a degli Studi di Napoli ``Federico II'', Complesso Universitario M. S. Angelo, via Cintia 21, 80126, Napoli, Italy}
\author{Domenico Ninno}
\affiliation{Dipartimento di Fisica ``E. Pancini'', Universit\`a degli Studi di Napoli ``Federico II'', Complesso Universitario M. S. Angelo, via Cintia 21, 80126, Napoli, Italy}
\affiliation{CNR-SPIN, c/o Complesso Universitario M. S. Angelo, via Cintia 21, 80126, Napoli, Italy}
\author{Giovanni Cantele}
\email{giovanni.cantele@spin.cnr.it}
\email{giovanni.cantele@unina.it}
\affiliation{CNR-SPIN, c/o Complesso Universitario M. S. Angelo, via Cintia 21, 80126, Napoli, Italy}

\date{\today}

\begin{abstract}
Twisted bilayer structures have emerged as a fascinating arena in condensed matter thanks to their highly tunable physics. 
The role of spin-orbit coupling (SOC) in twisted bilayers has gained increasing attention due to its potential for spintronics. 
Thus, it is appealing to propose new materials for constructing twisted bilayers with substantial SOC. 
In this work, the intriguing effects induced by twisting two layers of two-dimensional bismuthene are unraveled from large-scale first-principles calculations.
We show that spin-orbit coupling significantly affects the electronic properties of twisted bilayer bismuthene, even more than in its untwisted counterpart.
We carefully investigate how the interplay between the spin-orbit coupling and the twist angle impacts the band structure and spin textures of twisted bilayer bismuthene.
We find that the twist angle can be deemed a control knob to switch from a small-gap  semiconductor to a metallic behavior.
Most crucially, the accurate analysis of the energy bands close to Fermi energy reveals a twist-tunable splitting in the mexican-hat shape of the bands that can otherwise be obtained only by applying enormous electric fields. Our predictions provide insight into innovative bismuth-based technologies for future spintronic devices.

\end{abstract}

\maketitle


\section{Introduction}

Spin-orbit coupling (SOC) effect is the primary player in the field of spintronics due to its inner connection with the spin degree of freedom. In order to achieve spin manipulation, it is compelling to control the SOC-driven effects such as Dresselhaus \cite{DRESSELHAUS1955} and Rashba \cite{RASHBA1960}, leading to quantum spin Hall effect, spin–orbit torque, topological insulators and and many others, for future two-dimensional (2D) electronics \cite{MANCHON2015}.

So far, the most efficient strategies for engineering SOC into 2D devices, rely on external electric and magnetic fields tuning \cite{LIU2020,CHEN2021}. The van der Waals nature of 2D materials offers a wide variety of control knobs such as strain, stacking, and number of layers, that have been extensively studied to produce electrical and magnetic tunability \cite{LEI2022,LIN2023}.

Among them, the twist angle between layers has sparked a high interest in the 2D materials community as a unique platform to control a wide range of physical phenomena. Several intriguing phenomena arise from the moir\'e pattern as a consequence of twist, such as quantum spin Hall effect \cite{LEE2011}, unconventional superconductivity and flat bands \cite{CAO2018}, correlated electronic phase, and spin-polarized phases \cite{CAO2020,WANG2020}. 
Thus, recently, twistronics has been proposed as a brand new way to obtain SOC modulation and enforcement in systems like graphene/transition metal dichalcogenides (TMDs) hetero bi- and tri-layers \cite{PEZO2022,INGLA2022,CSABA2022} and TMDs/CrI$_{3}$ \cite{ZOLLNER2023}.

However, these promising results depend on proximity effect induced by the different nature of layers composing the heterostructures. Twisted group-IV buckled atomic layer have been proposed to obtain current-induced spin polarization (CISP) \cite{KITAGAWA2023}. Nevertheless, the search for a homostructure with a high SOC to provide spin order control via twisting is still ongoing. 

Bismuth bulk is the heaviest group-V semimetal. Being a quasi-layered crystal composed of buckled layers, it can be easily cleaved or grown along the (111) crystalline plane \cite{HOFMANN2006}. 
The Bi(111) surface displays topologically nontrivial electronic properties \cite{YAO2016} associated with the presence of metallic surface states \cite{KOROTEEV2004} that can be tuned by strain \cite{HIRAHARA2012} and thickness \cite{CANTELE2017}.

Bismuthene is the 2D bismuth single layer (1L) obtained from the Bi (111) surface. This compound has a huge SOC effect leading to Rashba split bands \cite{REIS2017,AKTURK2016,SUN2022}.
In bilayer bismuthene, inversion symmetry is restored by stacking two layers, while SOC shapes the bands in a mexican hat (MH)-like dispersion. 
Rashba and MH dispersion can be tuned in bismuthene by means of strain, electric field \cite{LIU2017} and sublattice half-oxidation (SHO) \cite{LIU2019}.
A moir\'e heterostructure of bismuthene/Black phosphorous has been shown modulating the topological edge states of bismuthene \cite{GOU2020}, while a study on bismuthene twisted homostructure is still missing.

In this work, by means of extensive first principles calculations, we unveil for the first time the electronic properties of twisted bilayer bismuthene (TB-Bi). We design the phase diagram of TB-Bi, predicting the twist-driven semiconductor to metal transition by considering a wide range of big twist angles. Most crucially, we find the possibility of tune SOC-driven MH parameters in twisted bilayer bismuthene. Interestingly, in addition to the flattening process, we find that the twist angle acts similarly to an external electric field, shaping the bands. Finally, we illustrate how spin textures evolve with twisting in comparison with the untwisted bilayer case, with and without the effect of an external electric field.
Our findings highlight the potential of twisted bilayer bismuthene as a candidate for future experimental realizations of twisted strong SOC materials for spintronics.

This paper is organized as follows: in Sec. II we describe the technical details of our first principles calculations.
In Sec. III we present geometrical and electronic properties of untwisted bilayer bismuthene. We discuss the modulation of the MH parameters with the inclusion of an external electric field.
In Sec. IV we present the geometrical and electronic properties of twisted bilayer bismuthene. We argue how the twist angle behaves similarly to an external electric field shaping the bands and the spin textures.
Finally, in Sec. V we present our conclusions.

\section{Methods}
\begin{figure*}[]
\includegraphics[width=0.8\linewidth]{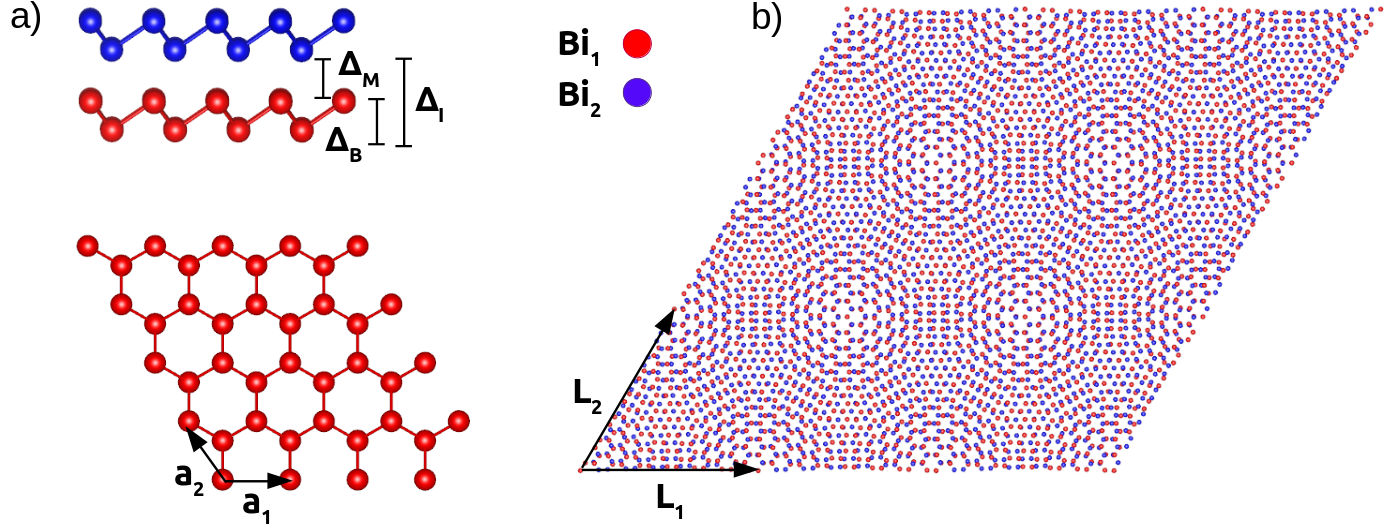}
    \caption{Geometrical properties of untwisted and twisted bilayer bismuthene. Red and blue atoms represent the first and the second layer respectively. Both structures are enlarged to a $3\times3\times1$ supercell to better highlight their pattern. a) Side and top view of AA stacked untwisted bilayer bismuthene. $\mathbf{a_{1}}$ and $\mathbf{a_{2}}$ are the in-plane lattice vectors defining the primitive unit cell, while the out-of-plane distances, as detailed in the text, are labelled as $\Delta_M$, $\Delta_B$ and $\Delta_l$. b) Moir\'e pattern generated by a twist angle of $5.0858$\degree corresponding to the  ($7$,$6$) supercell. The moir\'e periodicity is described by the lattice vectors $\mathbf{L_{1}}$ and $\mathbf{L_{2}}$.}
    \label{fig:1}
\end{figure*}
We perform first principles calculations in the framework of density functional theory (DFT), as implemented in the Vienna Ab Initio Package (VASP) \cite{VASP}, version 6.02.

We select two different projected augmented wave (PAW) pseudopotentials,  in the generalized gradient approximation (GGA) based on the Perdew-Burke-Ernzerhof (PBE) scheme. 
The first one contains only the $5$ Bi valence electrons ($6s^{2}6p^{3}$), while the second one comprises $15$ electrons, taking into account also the last Bi closed ($5d^{10}$) shell.
The kinetic energy cutoff for the plane-wave basis set is set to be 
$110$ and $250$ eV respectively. We include the closed ($5d^{10}$) shell in the valence electrons during the geometry optimization step in every structure under consideration, while band structure and density of states (DOS) calculations are carried out using only the $6s^{2}6p^{3}$ electrons. Indeed, as shown in the Supplemental Material (SM)~\cite{supp}, neglecting the $5d^{10}$ levels, that lie deep below $-20$ eV, results in no significant change in electronic electronic properties around the Fermi energy, which are dominated by the $p$-orbital character. This procedure allows us to include SOC in the electronic properties, even for lower twisted angles which would be otherwise computationally very demanding.

In order to sample the Fermi surfaces of the structures, a Gaussian smearing of 0.02 eV is applied.
The Brillouin zone of the untwisted films is sampled by $8\times8\times1$ Monkhorst-Pack $k$-point grid, while the one of the twisted structures is scaled according to the dimension of the supercell to keep the sampling accuracy nearly constant. 
Considering the $z$-axis as the stacking direction for the structure, we keep the supercell size along that direction fixed for all the analyzed structures. 
A $\sim14$ {\AA} vacuum space is added along the out-of-plane direction, proven to be sufficient to prevent periodic replicas from interacting with each other.
For all the considered structures, the atomic positions are optimized using the conjugate gradient algorithm in the PBE \cite{PBE} scheme, with a convergence criterion such that the total energy difference between consecutive structural optimization steps is less than $10^{-4}$ eV and all components of the forces acting on the atoms must be less than $10^{-3}$ eV/{\AA}.

In order to take into account van der Waals interactions between layers, non local Hamada's rev-vdW-DF2 functional \cite{rev-DF2-vdW} is employed in the geometry optimization step in both bilayer and twisted bilayer structures.

\section{Untwisted Bilayer Bismuthene and the effect of the electric field }
\begin{figure*}[]
\centering
\includegraphics[width=1.0\textwidth]{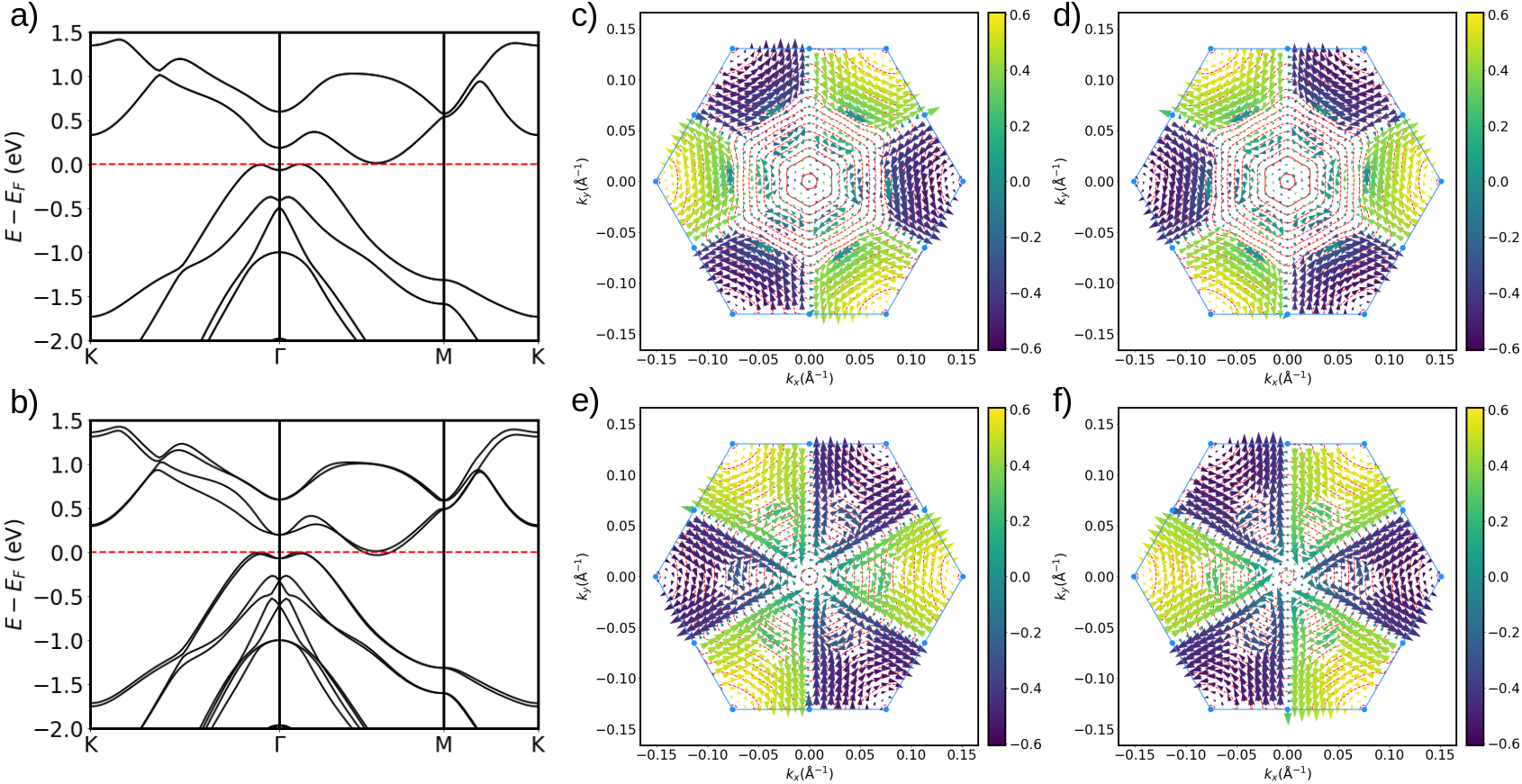}
\caption{(a) Band structure of untwisted 2L bismuthene along the K-$\Gamma$-M-K path in the two-dimensional hexagonal first Brillouin zone (BZ). 
Red dashed line corresponds to the Fermi energy. (b) Same as in (a) under the effect of an external electric field of $0.6$ eV/\AA. 
(c-d) Spin textures of the two topmost valence bands shown in (a). (e-f) Spin textures of the two topmost valence bands shown in (b). In (c-f) the color bar corresponds to the expectation value of $\hat{S}_{z}$ (yellow spin up and violet spin down). The direction of the arrows corresponds to the in-plane components of the spin, while the magnitude of the arrows is proportional to $\sqrt{S_{x}^{2}+S_{y}^{2}}$. Red lines correspond to constant energy contours, blue lines highlight the first BZ, blue dots
highlight K and M points at the boundary of the first BZ.}
\label{fig:2}
\end{figure*}
\begin{figure*}[]
\includegraphics[width=1.0\linewidth]{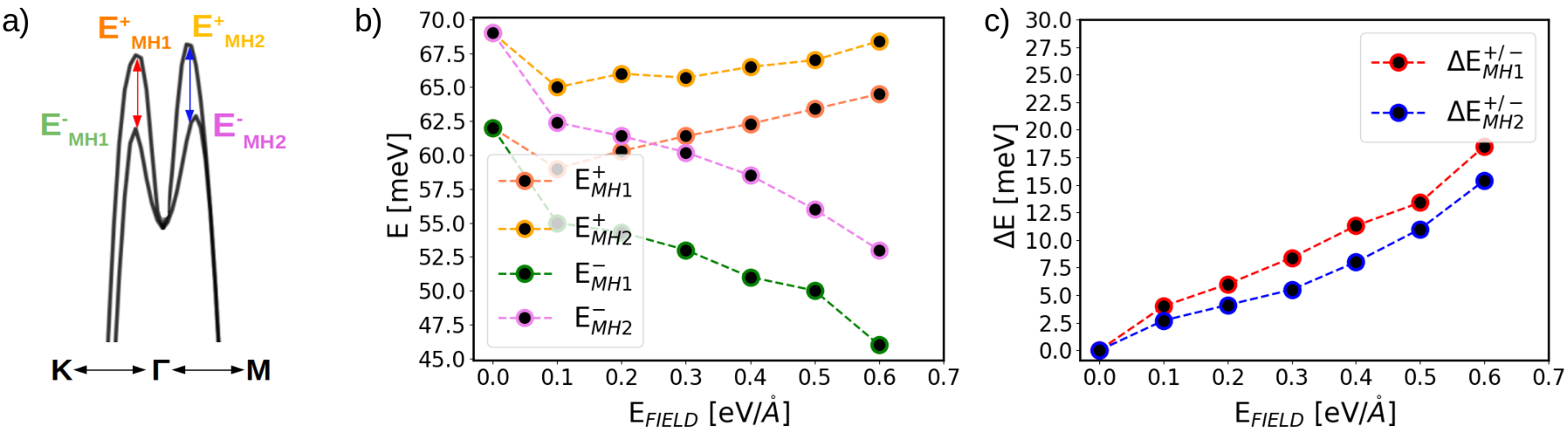}
    \caption{Evolution of the MH parameters under an external electric field (ranging $0.1$-$0.6$ eV/\AA). (a) Sketch of the anisotropic MH. The four edges energies are depicted, labeled as E$_{MH1}^{+/-}$ and E$_{MH2}^{+/-}$, where the superscripts 
    $+$ and $-$ define upper and lower peaks (associated to the two different valence bands) within the same path,
    respectively, and the subscripts 1 and 2 identify the peaks along the $\Gamma-K$ and $\Gamma-M$ paths, respectively. 
    (b) MH edge energies as a function of the external electric field. (c) MH splitting of each peak 
    $\Delta E^{+/-}_{MH1,2} = \left|E^{+}_\mathrm{MH1,2} - E^{-}_\mathrm{MH1,2}\right|$ as a
    function of the external field.}
    \label{fig:3}
\end{figure*}
We start discussing the properties of untwisted bilayer (2L) bismuthene.
Let us preliminary notice that in a material, as time reversal symmetry holds ($E(k,\uparrow) = E(-k,\downarrow)$), any energy band is degenerate with respect to a simultaneous flipping of $k$-vector and spin.
Moreover, if also the inversion symmetry holds, as it happens in the case of a 2L, an additional degeneracy with respect of $k$-vector flipping is present at fixed spin component, i.e. $E(k,\uparrow\downarrow) =  E(-k,\uparrow\downarrow)$. 
In other words, the combination of the time reversal and inversion symmetry leads to the Kramer's spin degeneracy, that is, $E(k,\uparrow) =  E(k,\downarrow)$. 
As a consequence, spin degenerate bands cannot be split in a 2L. 

In the case of huge SOC 2D materials with inversion symmetry breaking, such as bismuthene, Rashba effect shows up in the monolayer limit. 
Conversely, bilayer bismuthene inherits inversion symmetry from the bulk, and thus the band structure assumes a MH shape. In this configuration, as we will discuss, each individual band is twice degenerate in terms of spin degree of freedom, even in the presence of SOC. 
It has been shown that this degeneracy can be lifted by application of an external electric field or with a surface charge transfer \cite{LIU2017}.

 Bilayer bismuthene consists in two buckled single layers stacked along the [111] rhombohedral crystallographic direction, with four atoms in the unit cell. Each monolayer possesses the peculiar 2D buckled honeycomb lattice of group-V pnictogens: as depicted in Fig.~\ref{fig:1}, the primitive cell is composed of two atoms alternately in an upper and a lower plane, the spacing between which is the buckling distance $\Delta_{B}$. The buckling derives from the weak $\pi-\pi$ bonding making advantageous the dehybridization of $sp^{2}$ planar bonding and the formation of $sp^{3}$-like hybridization. It must be pointed out that the buckling conformation is inherited by the bulk counterpart, as the freestanding bismuthene stabilizes by buckling. 
 Beside that, during the synthetization procedure, monolayer bismuthene becomes flat as a result of the interaction with the substrate, whose effect is not considered in this work \cite{REIS2017,SUN2022}. In the 2L we can define two more distances $\Delta_{M}$ and $\Delta_{I}$ (see Fig. \ref{fig:1}) which respectively correspond to the minimum distance along the $z$-axis between two Bi atoms and the interlayer distance ($\Delta_{I}=\Delta_{B}+\Delta_{M}$) (the optimized geometrical parameters are reported in the SM~\cite{supp}). 

Previous studies have demonstrated that the stacking influences the dynamic stability of 2L bismuthene \cite{AKTURK2016}, and thus we choose to focus on the most stable configuration, the AA stacking. As detailed later, in order to build the twisted bilayer, we started from the AA stacking in which each Bi atoms of the first layer have the same in-plane coordinates of the Bi atoms of the second layer (Fig.~\ref{fig:1}).

To gain insight into the effect of the twist angle in the electronic properties of bismuthene, we started from the untwisted bilayer, evaluating the band structure with and without the application of an external electric field.
The band structure of untwisted 2L bismuthene along the $K-\Gamma-M-K$ path in the two-dimensional first Brillouin zone (BZ) is presented in Fig.~\ref{fig:2} in the PBE+SOC framework.
Untwisted 2L is a perfectly compensated semi-metal: the Fermi level crosses both the valence band maxima (VBMs) peaks situated along the $K-\Gamma$ and $\Gamma-M$ paths while the conduction band maximum (CBM) is along the $\Gamma-M$ path.

As mentioned above, untwisted 2L bismuthene is characterized by a MH shape in the valence bands. We label the characteristic parameter of the MH dispersion in the valence band as the energy E$_{MH}$, corresponding to the 
energy difference between each peak and the valence band energy at $\Gamma$ point. Similarly, k$_{MH}$ labels
the distance in reciprocal space of each peak from $\Gamma$ point. 
It turns out that, as already reported in the literature \cite{WICKRAMARATNE2014}, 
2L bismuthene shows an anisotropy in the MH shape, since the heights of the peaks are slightly different in energy along the K-$\Gamma$ and the $\Gamma$-M paths of the first BZ. 
As depicted in Fig.~\ref{fig:3}, we take into account the anisotropy by defining two different energies, namely E$_{MH1}$ and E$_{MH2}$, where the subscripts 1 and 2 refer to the peaks along the K-$\Gamma$ and the $\Gamma$-M paths, respectively. Moreover, for each peak, a superscript
$+$ or $-$ is needed to label the upper and the lower valence bands, respectively, in the cases
in which their degeneracy is removed. For example, $E_{MH1}^+$
will refer to the energy of the upper valence band at the first peak (along the K-$\Gamma$ segment).
 
For the untwisted bilayer, the $\Gamma$-M peak is slightly higher then the K-$\Gamma$ one,  namely $E_{MH1}=62$ meV and $E_{MH2}=69$ meV (the two valence bands are degenerate),  while the radius of the MH is k$_{MH}=0.01599$ {\AA}$^{-1}$.

By introducing an external electric field, the twofold degeneracy of the band of the untwisted bilayer bismuthene is lifted, as shown in Fig.~\ref{fig:2}(b) which leads to band splitting. We label the four MH edge energies as E$_{MH1}^{+/-}$ and E$_{MH2}^{+/-}$, the apex $+/-$ defining upper/lower peaks within the same path, while $\Delta E^{+/-}$ is the splitting introduced by the external field of each peak (Fig.~\ref{fig:3}(a)). 

We study the evolution of the MH parameters with the modulation of the external electric field for a range of $0.1$-$0.6$ eV/\AA. We observe a linear increase of the splitting for both peaks with the field intensity (Fig.~\ref{fig:3}(b)).
However, the magnitude of the splitting is different for the two peaks as $\Delta E_{MH1}^{+/-}>\Delta E_{MH2}^{+/-}$, for each value of the electric field (Fig.~\ref{fig:3}(c)). In other words, the applied electric field enhances the anisotropy of the peaks.  
The spin-split bands along the K-$\Gamma$ path are shrunk by a tiny factor of $6\times10^{-4}$ {\AA$^{-1}$} in $k$ space with respect to the initial $k_{MH}$ value. 
These results are in good agreement with Liu and coworkers \cite{LIU2017}, and represent the starting point to study the effect of twist, as we will see in the next section.

Spin textures in bismuthene are significant due to their relationship with the material topological properties and its prospective use in spin-based electronics. 
As an example, a recent work found that adjusting bulk-edge interactions in planar bismuthene results in persistent edge spin textures, even under heavy external electric fields \cite{YIN2023}.
In Fig.~\ref{fig:2}(c-d) we report the calculated spin textures of the two topmost valence bands in the case of 
untwisted bilayer Bi. The modifications of these spin textures under the effect of an external electric field 
of 0.6 eV/{\AA} is reported in Fig.~\ref{fig:2}(e-f).

As we can see from the figure, the spins in the MH shaped bands are bound to circulate in the in-plane directions in most of the BZ as a result of the strong spin-orbit coupling, while around $K$ and $K^{'}$ the spin are purely out-of-plane.
The spin textures in the zero field case show a vortex-like pattern 
around $K$ and $K^\prime$ with alternating helicity, whereas an hexagonal pattern shows up around $\Gamma$. 
Electric field can be employed to tune the spin textures upon switching the electric polarization \cite{LIUPIC2019}. Indeed, by switching on an electric field of $0.6$ eV/\AA, the spin textures are modified as follows.
A swap in helicity occurs between the two topmost valence bands, together with the formation of triangular slices inside the $M-\Gamma-M^{'}$ region of the BZ, each one with a small hole resembling a petal, forming a flower-shape around $\Gamma$. 

In both cases, the two topmost valence bands display the same spin textures, with flipped spin, and thus no exotic phenomena occurs. This picture is modified by twisting as reported in the next section.

\section{Twisted Bilayer Bismuthene}
After characterizing the untwisted bilayer, we study twisted bilayer bismuthene (TB-Bi) systems up to six different twist angles. The method to build a commensurate cell starting from two equal AA-stacked layers of hexagonal bismuthene lattices is the following. We rotate around an axis orthogonal to the layers and passing through
two Bi atoms, one on top of the other, belonging to the two planes (that therefore preserve their initial AA stacking).
\begin{figure*}[]
\includegraphics[width=1.0\linewidth]{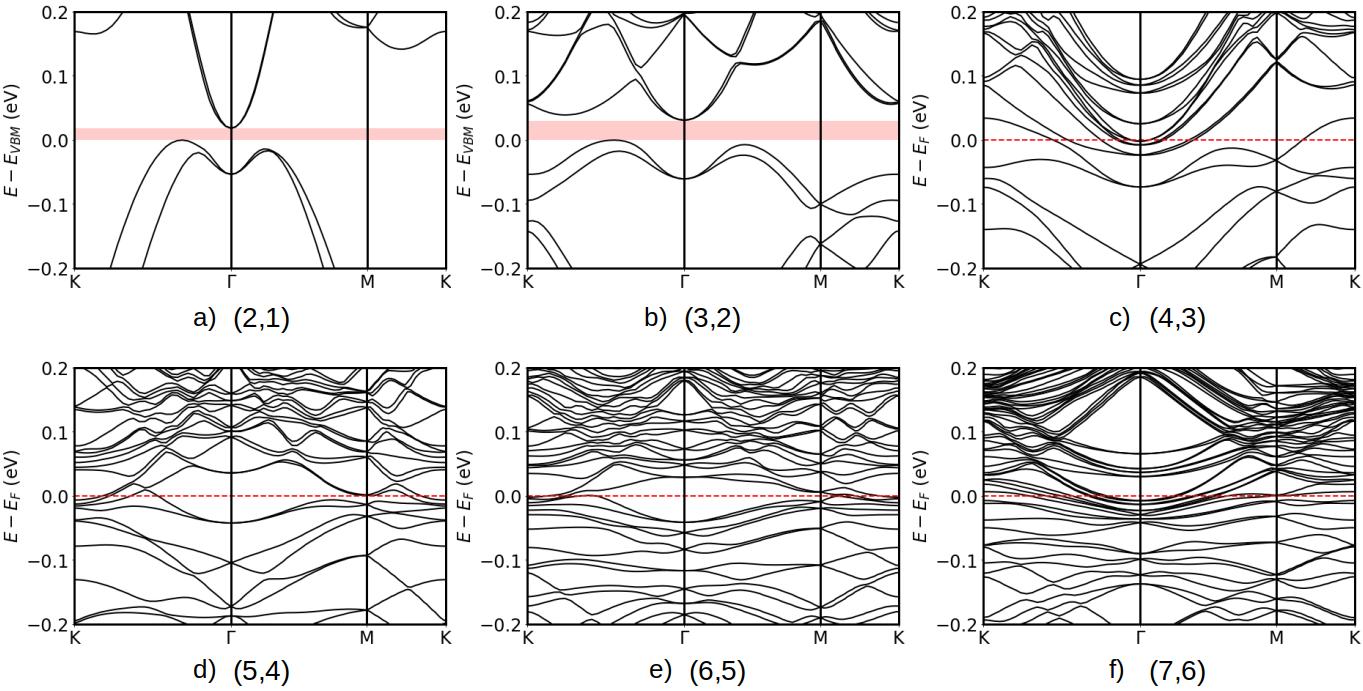}
    \caption{Twist-tunable band structure of TB-Bi along the K-$\Gamma$-M-K path in the two-dimensional hexagonal first mini BZ
    (mBZ). The red box corresponds to the gap for insulating systems, while the red dashed line corresponds to the Fermi energy for metallic systems.}
    \label{fig:4}
\end{figure*}
As shown in literature \cite{KOSHINO2012}, we can define the moir\'e lattice primitive vectors $\mathbf{L_{1}}$ and $\mathbf{L_{2}}$ starting from the lattices of the two layers as:
\begin{equation*}
    \mathbf{L_{1}}=n\mathbf{a_{1}^{(1)}}+m\mathbf{a_{2}^{(1)}}=n'\mathbf{a_{1}^{(2)}}+m'\mathbf{a_{2}^{(2)}}
\end{equation*}
where $\mathbf{a_{1,2}^{(i)}}$ with $i=1,2$ are the primitive lattice vectors of the first and the second bismuthene layers respectively. $\mathbf{L_{2}}$ is then obtained by rotating $\mathbf{L_{1}}$ by $60\degree$. The two pairs of integer indices $(n,m)$ and $(n',m')$ can be set to be equal, so that the TB pattern is specified by a single pair of moir\'e indices $(n,m)$. The twist angle $\theta$ can be derived as a function of the latter as:
\begin{equation*}
    \cos{\theta}=\frac{1}{2}\frac{n^{2}+m^{2}+4nm}{n^{2}+m^{2}+nm},
\end{equation*}
whereas the lattice constant $L=|\mathbf{L_{1}}|=|\mathbf{L_{2}}|$ is:
\begin{equation*}
    L=a\sqrt{n^{2}+m^{2}+nm}=a\frac{|n-m|}{2\sin{\theta/2}}    
\end{equation*}
and thus, the length of the moir\'e cell $L$ is inversely proportional to the twist angle $\theta$. The moir\'e pattern of the lowest twist angle in this work is depicted in Fig.~\ref{fig:1}(b).
A summary of the moir\'e parameters of the system under consideration is summarized in Table \ref{tab:2Moire}. 

\begin{table}[ht!]
    \centering
    {
    \begin{tabular}{c|c|c|c}
     $(n,m)$  & $\theta$ (\degree) &  $N$ & $a$ ({\AA})\\
    \hline
    (2,1) &21.7868 &28 &11.6942\\
    (3,2) &13.1736 &76 &19.2663\\
    (4,3) &9.4300 &148 &26.8858\\  
    (5,4) &7.3410 &244 &34.5213\\    
    (6,5) &6.0090 &364 &42.1641\\    
    (7,6) &5.0858 &508 &49.8109  
    \end{tabular}
    }
    \caption{Calculated geometric parameters for TB-Bi. The integers $n$ and $m$
    are the moir\'e indices, $\theta$ is the twist angle, $N$ is the number of atoms in the primitive cell, $a$ is the supercell lattice parameter.}
    \label{tab:2Moire}
\end{table}
The band structures of TB-Bi for the six considered twist angles are shown in Fig.~\ref{fig:4} along the $K-\Gamma-M-K$ path of the mini-BZ in the PBE+SOC scheme. 
At first glance, we can immediately infer that a semimetal to a narrow band-gap semiconductor transition occurs. Indeed, going from the untwisted bilayer, in which Fermi level intersects both the VBM and the CBM, to the first twist angle (namely $(2,1)$ supercell, $\theta=21.7868$\degree) a small indirect gap of $E_{g}=$18 meV opens up, with the VBM located along the
$K-\Gamma$ line and the CBM located at $\Gamma$. 
The next twist angle, $\theta=13.1736\degree$, corresponding to the $(3,2)$ supercell, follows the trend, displaying a semiconductor behaviour with an indirect band gap increased to $E_{g}=30$ meV. 
Based on these results, we can conclude that the large-twist angle systems show a twist-driven indirect band-gap tunability, together with the typical band flattening mechanism that occurs in twisted bilayers. 
This becomes evident comparing the band dispersion of the two topmost valence bands of panels (a) and (b) in Fig.~\ref{fig:4}. 

An abrupt inversion of the trend then occurs when switching to the third supercell $(4,3)$ with twist angle $\theta=9.4300$\degree. 
There is a crossover between valence and conduction bands that are now overlapped. Moreover, the two last flat valence bands greatly split: the higher-in-energy valence band crosses and exceeds the Fermi level along the $K-\Gamma$ and $M-K$ path, while the lower-in-energy valence band remains below the Fermi level for every k-points. 
We can infer that the third twist angle is the critical one, from which onwards twisted bilayer bismuthene displays metallic behaviour. Indeed, the next three twist angles, corresponding to the supercells $(5,4)$, $(6,5)$ and $(7,6)$ are metallic.

\begin{figure*}[tb]
\includegraphics[width=1.0\linewidth]{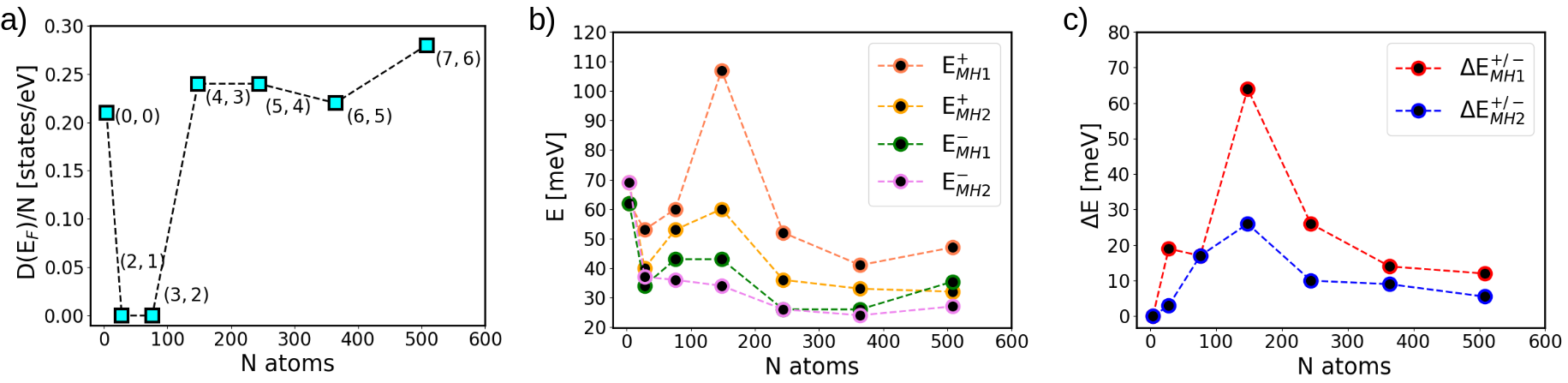}
    \caption{(a) Twist-tunable phase diagram of TB-Bi. Squares corresponds to the DOS per atom at the Fermi level $D(E_{F})/N$ (with $N$ number of atoms in the twisted supercell). (b-c) Evolution of the MH parameters, as defined in Fig.~\ref{fig:3} with twisting. The MH edge energies as a function of $N$ are shown in (b), whereas the splitting of each peak $\Delta E^{+/-}_{MH1,2}$ introduced by the twist angle is depicted in (c).}
    \label{fig:5}
\end{figure*}

We can visualize the results by building a twist-tunable phase diagram. This can be done by plotting the density of states (DOS) per unit cell at the Fermi level $D(E_{F})/N_{Atoms}$ as shown in Fig.~\ref{fig:5}(a). 
The DOS at the Fermi level is zero for the first two angles, while a peak appears in the $(4,3)$ supercell. 
A similar metallic behaviour shows up for the other twisted bilayers, with smaller twist angles, with the
DOS per atom increasing as the twist angle decreases. As confirmed by comparing the band structures in  Fig.~\ref{fig:5},
this corresponds to the filling of the bottom conduction bands, and should results in an enhanced electrical conductivity as the twist angle decreases.

Another remarkable feature of the band structures of Fig.~\ref{fig:4} is that a band flattening mechanism occurs 
upon twisting. This concerns, in particular, the two topmost (SOC-split) valence bands. 
However, these two bands maintain the SOC-driven MH shape in the $K-\Gamma$ and $M-K$ paths. The emergence of flat bands appears at small twist
angles in twisted bilayer graphene, and it is ascribed to the combined action of moir\'e superlattices and interlayer interactions \cite{CAO2018,CAO2020,CANTELE2020,PhysRevB.99.155429,PhysRevB.99.195419}. 
However, it has been demonstrated that flat bands can also form at twist angles greater than a few degrees, for example in twisted bilayer hexagonal boron nitride \cite{ZHAO2020}, and two-dimensional indium selenide \cite{TAO2022}. Furthermore, a recent study demonstrated that it is possible to realize continuously tunable superflat bands due to the presence of localized states around the AA stacking regions in general twisted bilayer systems \cite{TAO2022}. 

The signature of the tunability of the electronic properties with twisting does not rely only on the semiconducting
or metallic behavior of the considered structures. To better assess this circumstance, let us recall that in untwisted $2$L bismuthene, in the absence of SOC, each individual band has a two-fold spin degeneracy. 
As the SOC is turned on, the degeneracy is lifted, as it is evident from the band splitting at all twisted angles
depicted in Fig.~\ref{fig:4}. Remarkably, the twofold degeneracy lifting arises without any external applied electric field. 
A similar mechanism occurs in few-layer Bi$_2$Se$_3$ and PtSe$_2$ heterostructures as a result of the band-alignment, with layer dependent spin-splittings \cite{SATTAR2020}.
In our case, we can associate this effect to the combination of a strong SOC and the breaking of the centrosymmetric geometry of 
the untwisted $2$L induced by the moir\'e superlattice. 
For systems that in the $2L$ form preserve a centrosymmetric geometry the band splitting in the presence of SOC can be induced only by an external perturbation,
such as an electric field.
In the present case, instead, twisting plays the role of a ``pseudo electric field'', making the two layers not equivalent. 
The analogy may be stretched even further since, as we will see in the following, the amount of splitting can be controlled with the electric field intensity, as well as with the twist angle.
To this purpose, we can compare the evolution of MH parameters as a function of the twist angle with the 
electric field-induced band splitting, previously discussed for the untwisted bilayer. 
Our results are shown in Fig.~\ref{fig:5}(b-c). We observe that for large twist angles (small supercells), the magnitude of the splitting tends to increase until it reaches a maximum. 
Again, we can define the angle $\theta=9.43^{\circ}$ as the critical angle after which we observe a change in the MH parameters trend. After that (that is, for smaller twist
angles and larger supercells) 
the previously discussed metallic behaviour is accompanied with a decrease of the splitting. 
A second point that is noteworthy to stress is that, as in the case of a bilayer under external electric field (compare with Fig.~\ref{fig:3}(c)), the electric field-induced change in the amplitude of the splitting is different for the two peaks.
Moreover, an ``enhanced'' anisotropy effect shows up when compared with the untwisted bilayer, in that $k_{MH1}\ne k_{MH2}$, that is, the two peaks arise at different distances from $\Gamma$ point. This can be attributed to a twist-induced anisotropy effect, because twisting causes both a rotation and a resizing of the BZ (compared to the untwisted one).
In the untwisted supercell, energy peaks along the $K-\Gamma$ path are lower in energy than those along the $\Gamma-M$ path. Twisted bilayers display the opposite behaviour.

Therefore, we can conclude that what mostly distinguishes the twist-induced changes of the MH peaks from the case of the untwisted bilayer under external electric field is the following.
As the BZ shrinks upon decreasing the twist angle, the positions of the peaks in $k$ space changes, with a consequent increasing of the MH anisotropy, as further detailed in the SM~\cite{supp}. 
Consequently, not only the twist angle acts as an electric field tuning the SOC-induced splittings, but also it can
be used as a control knob to tune anisotropy in the $k$ space. 

It must be pointed out that in their work Scammel and Scheurer \cite{SCAMMEL2023} show that SOC-split 
bands can be induced by a displacement field that breaks inversion symmetry in mirror-symmetric twisted trilayer graphene heterostructure with TMDs. Here, no external parameter is included, since SOC splitting arises from twisting the heavy Bi atoms.
\begin{figure*}[]
\includegraphics[width=1.0\linewidth]{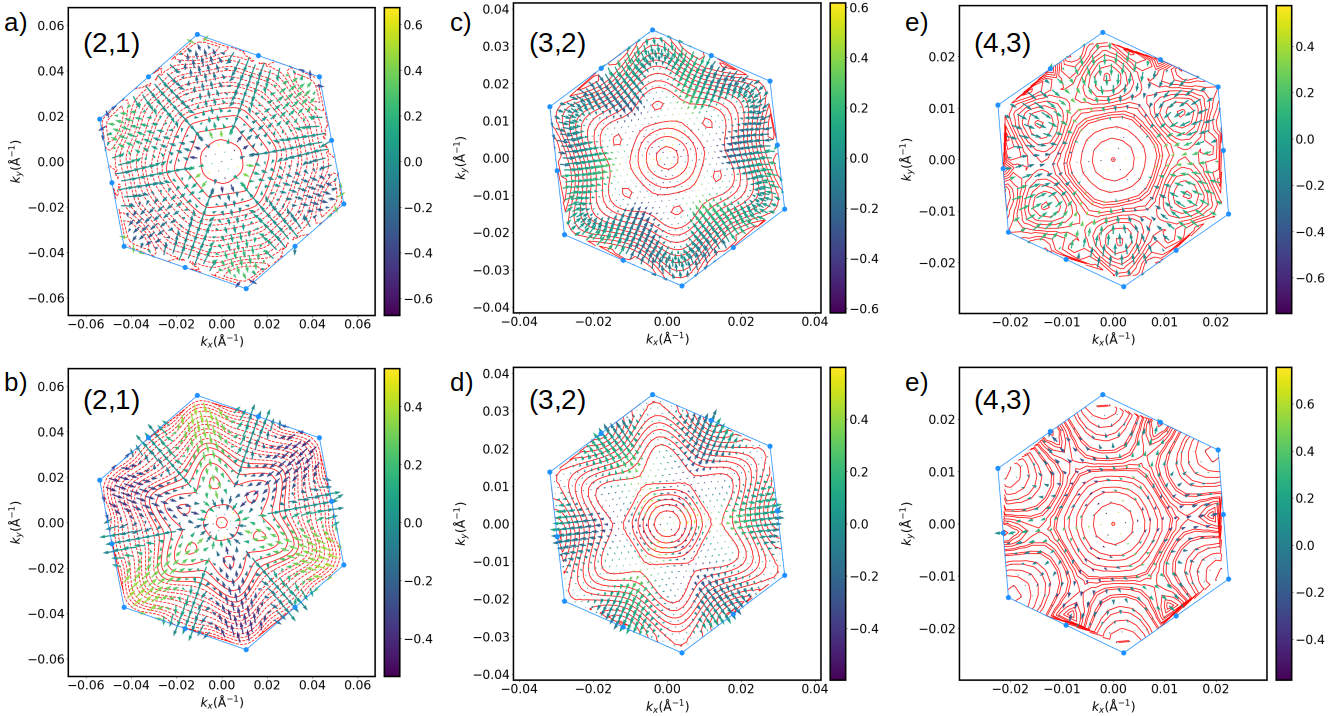}
    \caption{Twist-tunable spin textures of TB-Bi in the two-dimensional hexagonal mBZ.  The results are shown for 
    the two topmost valence bands of (a-b) ($2$,$1$), (c-d) ($3$,$2$) and (e-f) ($4$,$3$) supercells, respectively. The color bar corresponds to the expectation value of $\hat{S}_{z}$ (yellow spin up and violet spin down). Direction of the arrows corresponds to the in-plane components of the spin, while the magnitude of the arrows is proportional to $\sqrt{S_{x}^{2}+S_{y}^{2}}$. 
    Red lines correspond to constant energy contours, blue lines highlight the first mBZ, blue dots highlight K and M
    points at the boundary of the first mBZ.}
    \label{fig:6}
\end{figure*}

We conclude showing the spin textures evolution in TB-Bi upon twisting. It is known in literature that in the case of twisted bilayer graphene, spin textures are strongly influenced by the twist chirality and angle, as well as the doping level, leading to peculiar shapes \cite{YANANOSE2021,SBOYCHAKOV2022}.
In Fig.~\ref{fig:6} we show the calculated spin textures for the first three twist angles of TB-Bi.
First of all, we can clearly see that the spin textures are highly twist-tunable, since the spin pattern dramatically changes with twisting.

Twisting not only changes spin texture, but also causes considerable anisotropy compared to electric field-induced splittings in untwisted bilayer.
Indeed, as Figs.\ref{fig:2}(c-f) reveal, the spin patterns in the previous case show a perfect symmetry between otherwise (that is, in the absence of the external field) degenerate bands. These bands exhibit the same spin texture in the whole BZ, with the only difference being reverting the direction of the spin at each $k$ point.
On the other hand, such a symmetry gets broken in the twisted bilayers, due to the discussed anisotropy of the valence bands. 
In particular, the patterns of Fig.~\ref{fig:6} clearly reveal that the paired bands do not exhibit spin textures that can be exactly 
obtained from each other by reverting the sign of the spin vector at each $k$-point.
To support such a circumstance, we can also notice that the ``asymmetry'' between the bands that would be paired in the untwisted bilayer changes with the twist angle. 
Therefore, we can speak about twist-tunable spin textures, with the twist offering, with respect to the electric field, the further possibility
of breaking the symmetry between the two valence bands.
This effect could be influenced by the flattening mechanism induced by the twist, and it is not achievable by the action of an external electric field.
This result shows that it is possible not only to tune the spin textures by twisting but also to select two different spin configurations within the same twist angle thanks to the MH modification induced by the flattening, a peculiar behaviour that could be exploited for spintronic devices.

\section{Conclusions}

By performing extensive first-principles electronic structure calculations based on DFT, we predicted 
the electronic and spin properties of twisted bilayer bismuthene (TB-Bi).
To allow a better assessment of the effect of twisting,  the untwisted bilayer properties have been carefully investigated, with a particular focus on the MH shape of the topmost valence 
bands induced by the SOC. 
The twofold spin degeneracy of these bands can be broken after the introduction of an external electric field.
We commented on the evolution of the MH parameters (peak energies and splittings) as a function of the field.
The analysis of TB-Bi reveals that twisting might act as a ``pseudo'' electric field, in that it breaks the symmetry of the topmost
valence bands. In addition, combined with SOC, the twist angle acts as a tool to tune and control the MH parameters.
Remarkably, twisting induces an anisotropy of the MH, that shows up both in the position in $k$ space of the peaks and in their energies. 
These results show that twist angle can be employed in this context as a control knob for the band splitting, without relaying on any external control, which is usually achievable only with the application of a huge electric field. 
We derived the phase diagram of TB-Bi by evaluating the evolution of DOS at the Fermi level. We showed that a semimetal to a narrow band-gap semiconductor transition occurs going from the untwisted case to the first two larger twisted angles.
We identified $\theta=9.43^\circ$ as the critical angle after which the flattening mechanism induces a metal behaviour for the smaller twist angles.  

Finally, we showed that it is possible to tune the spin textures of bismuthene by twisting. We unveil an exotic behaviour of the spin patterns induced by the asymmetry of the MH shape, as a result of the twisted-induced flattening mechanism.  
This work propose bismuthene as a new candidate for twistronics, being a system with potentially high built-in spin control for future spintronic devices.

\section*{Acknowledgments}

We acknowledge the CINECA award under the ISCRA initiative, for the availability of high-performance computing resources and support.
Financial support and computational resources from MUR, PON "Ricerca e Innovazione 2014-2020", under Grant No. PIR01\_00011 (I.Bi.S.Co.)
are acknowledged.

\end{document}